\documentclass{pasj00}
\draft

\begin{document}
\SetRunningHead{T. Kuwabara, A. Taruya, and Y. Suto}
{Pairwise velocity distribution function}
\Received{2001 November 28}
\Accepted{2002 May 2}

\title{Modeling Pairwise Peculiar Velocity Distribution Function 
of Dark Matter from Halo Density Profiles}

\author{Takeshi \textsc{Kuwabara},\altaffilmark{1}
        Atsushi \textsc{Taruya},\altaffilmark{1,2}
        and Yasushi \textsc{Suto}\altaffilmark{1,2}}
\altaffiltext{1}{Department of Physics, School of Science, 
                 University of Tokyo, Tokyo 113-0033}
\altaffiltext{2}{Research Center for the Early Universe (RESCEU),\\
                 School of Science, University of Tokyo, Tokyo 113-0033}
\email{kuwabara@utap.phys.s.u-tokyo.ac.jp,
       ataruya@phys.s.u-tokyo.ac.jp, \\
       suto@phys.s.u-tokyo.ac.jp}
\maketitle
%
\begin{abstract}
We derive the pairwise peculiar velocity distribution function of dark
matter particles applying the dark matter halo approach. Unlike the
previous work, we do not assume a Gaussian velocity distribution
function of dark matter in a single halo, but compute it
self-consistently with the assumed density profile for dark matter halo.
The resulting distribution function is well approximated by an
exponential distribution which is consistent with the previous
observational, numerical and theoretical results.  We also compute the
pairwise peculiar velocity dispersion for different density profiles,
and provide a practical fitting formula.  We apply an empirical biasing
scheme into our model and present prediction for pairwise peculiar
velocity dispersion of {\it galaxies}, and reproduce the previous
results of simulations using our semi-analytical method.
\end{abstract}
%
%
\section{Introduction}

Since Davis \& Peebles (1983) first analyzed the anisotropy in the
galaxy distribution from Center for Astrophysics (CfA) redshift catalog,
it has been recognized that the peculiar velocity field of galaxies
induces a significant systematic effect in statistics of observed galaxy
distributions in redshift space. In particular, virialized random motion
of galaxies produces an elongated pattern of galaxy distribution along
the line of sight, called {\it finger-of-God}.  This effect
significantly suppresses the amplitude of the two-point correlation
function of galaxies in redshift space, especially on scales below $\sim
3h^{-1}$Mpc.

The proper account of this redshift-space distortion requires a detailed
model for the pairwise velocity distribution function (hereafter, PVDF)
of galaxies. Davis \& Peebles (1983) discovered that the PVDF of the CfA
galaxy sample is approximately described by an exponential distribution,
instead of a Gaussian:
\begin{equation}
 f_{12}(v_{12};r_{12}) = \frac{1}{\sqrt{2}\sigma_{12}(r_{12})}
                         \exp\left(-\frac{\sqrt{2}v_{12}}{
                            \sigma_{12}(r_{12})}\right),
\label{eq:exp}
\end{equation}
  where $r_{12}$ is the separation length and $v_{12}$ denote
  the pairwise peculiar velocity between a pair of galaxy along the
  line-of-sight direction.
  The quantity
 $\sigma_{12}(r_{12})$ is
the peculiar velocity dispersion (PVD).  The exponential form of PVDF
was confirmed also later by analyses of $N$-body simulations of dark
matter particles and of other samples of galaxies (e.g., Efstathiou et
al. 1988, Fisher et al. 1994, Marzke et al. 1995).

Theoretical models for the origin of the exponential PVDF of dark matter
were put forward by Sheth (1996) and Diaferio \& Geller (1996), and more
recently by Sheth \& Diaferio (2001). They phenomenologically introduced
a nonlinear model of PVDF using the Press--Schechter formalism. To be
more specific, they assume that any dark matter particle belongs to one
of virialized clumps (dark halos) with the 1-point velocity distribution
function being a Maxwellian form.  If one considers sufficiently small
scales, the particle pairs of those separations are likely to be in the
same halo, and then their PVDF is approximately given by
\begin{equation}
 f_{12}(v_{12};r_{12})=
 \frac{\displaystyle 
\int dM \, n(M) \, N_{\rm pair}(r_{12}|M) \, f_{12,\rm 1h}(v_{12}|M)}
{\displaystyle \int dM \, n(M) \, N_{\rm pair}(r_{12}|M)},
\label{eq:shethp}
\end{equation}                       
where $n(M)$ is the mass function of dark halos, $f_{12,\rm
 1h}(v_{12}|M)$ is the PVDF of dark matter particles within a halo of
 mass $M$. The quantity $N_{\rm pair}(r_{12}|M)$ represents the
 statistical weight proportional to the number of particle pairs with
 separation $r_{12}$ in the halo:
\begin{equation}
 N_{\rm pair}(r_{12}|M)=\int d\boldsymbol{r}^3_1
                        \int d\boldsymbol{r}^3_2
                        \rho(\boldsymbol{r}_1|M)
                        \rho(\boldsymbol{r}_2|M)
              \delta_D(r_{12}-|\boldsymbol{r}_1-\boldsymbol{r}_2|) ,
\label{eq:pairn}
\end{equation}
where $\rho(r|M)$ is the density profile of the halo of mass $M$, and
$\delta_D$ is the Dirac delta function.  Adopting the singular
isothermal distribution as a particular choice of the dark halo profile,
Sheth (1996)  showed that the scale-free
model of $P(k) \propto k^n$ with $n=-1$ exactly reproduces the
exponential PVDF (\ref{eq:exp}).

While a perturbation theory (Seto \& Yokoyama 1998; Juszkiewicz et
 al. 1998) also qualitatively explained why the Gaussian initial models
 approach the exponential PVDF, the above model is much more successful
 quantitatively.
   Further, a significant influence of the finger-of-Got effect appears at
   small scale, where the perturbative approach cannot be applied.
 Therefore in the present paper, we attempt to improve
 the Sheth (1996) model (\ref{eq:shethp}) for the PVDF in several
 aspects; first, we consider more popular CDM models instead of the
 scale-free power-spectra. Second, we adopt a series of more realistic
 density profiles for dark halos (Hernquist 1990; Navarro, Frenk \&
 White 1997; Fukushige \& Makino 1997, 2001a; Moore et al. 1998; Jing \&
 Suto 2000).  Third, we derive the one-point PVDF of dark matter
 particles in a halo directly from the Abel integral of the above
 density profiles, instead of assuming the Maxwellian form {\it a
 priori}. This approach is important since we can incorporate the scale-
 and mass-dependence of the PVD in a consistent fashion unlike the
 previous modeling. Finally we also apply the selection function
 following Jing, B\"orner \& Suto (2002) so as to phenomenologically
 attempt to predict the PVD of {\it galaxies} out of that of dark matter
 particles.

 This paper is organized as follows; Section 2 describes our improved
 modeling for the PVDF on the basis of the dark matter halo approach. In
 \S 3, we present the resultant PVD in various cosmological
 models and discuss how the underlying halo profiles are sensitive to
 those results.  We also provide a simple fitting formula of the PVD in
 the currently popular spatially-flat CDM model, which is useful in
 modeling the redshift-distortion effect. Then we attempt to consider
 the effect of the spatial biasing of galaxies relative to the dark
 matter particles on the PVD by applying a phenomenological biasing
 scheme.  Finally section 4 is devoted to summary and
 conclusions.

\section{A Dark Matter Halo Approach to Compute the
Pairwise Velocity Distribution Function}

\begin{figure}
  \begin{center}
  \FigureFile(160mm,100mm){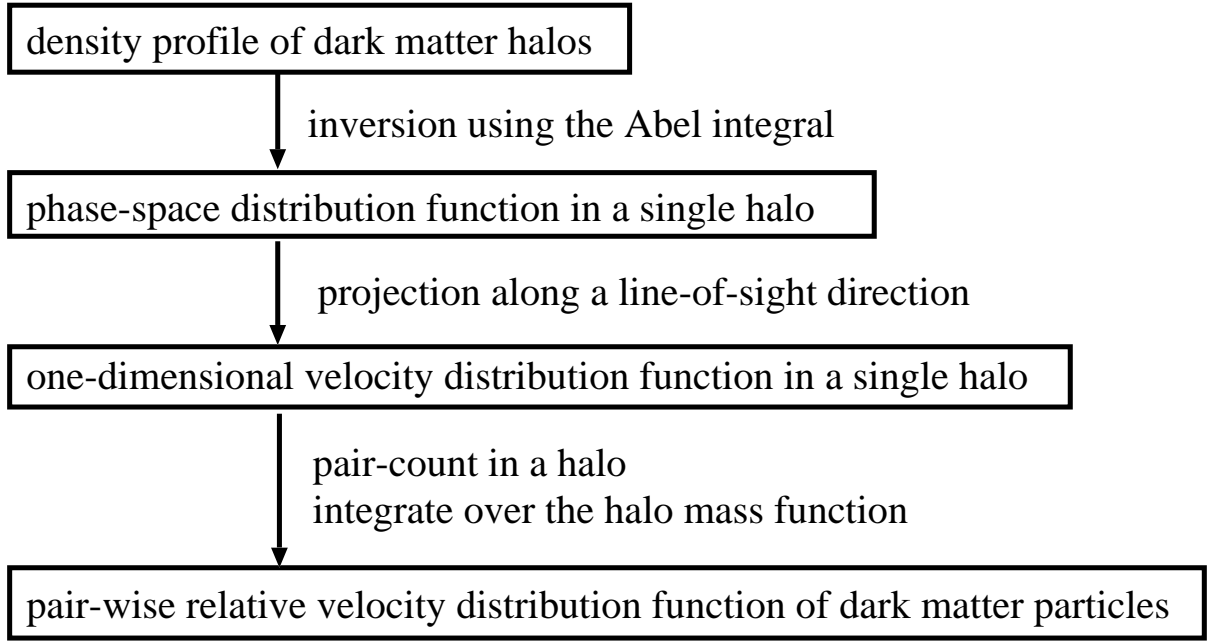}
  \end{center}
  \caption{Schematic illustration of our procedure to compute the
pairwise velocity distribution function.} \label{fig:pvdf_procedure}
\end{figure}

Our present method to compute the PVDF is schematically shown in Figure
\ref{fig:pvdf_procedure}. We will describe the details of the procedure
below. Throughout the paper, we consider the three representative CDM
models parameterized by the density parameter $\Omega_0$, the
dimensionless cosmological parameter $\lambda_0$, the amplitude of the
mass fluctuation smoothed over the top-hat radius of $8h^{-1}$Mpc,
$\sigma_8$, and the Hubble constant in units of 100 km/s/Mpc $h$;
standard CDM ($\Omega_0=1.0,\lambda_0=0,\sigma_8=0.6,h=0.5$; SCDM),
lambda CDM ($\Omega_0=0.3,\lambda_0=0.7,\sigma_8=1.0,h=0.7$; LCDM), and
open CDM ($\Omega_0=0.45,\lambda_0=0,\sigma_8=0.83,h=0.7$; OCDM). Those
models are normalized to satisfy the X-ray cluster abundances (Kitayama
\& Suto 1997).

\subsection{Density profile} \label{subsec:rho}

The density profile of dark matter halos plays a key role in our
method. Following the recent suggestions from high-resolution
numerical simulations, we adopt
the following specific form:
\begin{equation}
\label{eq:univ}
 \rho(r|M)=
 \left\{\begin{array}{cc}
  \displaystyle
  \frac{\bar\rho(z) \, \delta_c}{
    (r/r_s)^\alpha
    (1+r/r_s)^\nu} & (r< r_{\rm vir}) \\
  \displaystyle
  0 & (r>r_{\rm vir})
 \end{array}\right. .
\end{equation}
In the above, $M$ is the mass of the halo,
 $\bar\rho(z) \equiv \Omega_0 \rho_{\rm c0} (1+z)^3$ is the mean
density of the universe at $z$, $\rho_{\rm c0}$ is the present
critical density, $\delta_\mathrm{c}(M)$ is the characteristic density
excess, and $r_{\rm vir}(M)$ and $r_\mathrm{s}(M)$ indicate the virial
radius and the scale radius of the halo, respectively.

The virial radius is defined according to the spherical collapse model as
\begin{equation}
 r_{\rm vir}(M) \equiv \left(\frac{3M}{
                          4\pi\bar{\rho} \Delta_{\rm nl}}\right)^{1/3} .
\label{eq: r_vir}
\end{equation}
We use the following expressions (Kitayama \& Suto 1996) for the
 critical over-density $\Delta_{\rm nl}$:
\begin{equation}
 \Delta_{\rm nl}(\Omega_0,\lambda_0) = \left\{
   \begin{array}{ll}
     \displaystyle
     18\pi^2(1+0.4093\omega_{\rm vir}^{0.9052})
              & (\Omega_0<1, ~ \Omega_0+\lambda_0=1) \\
     \displaystyle
     4\pi^2\frac{(\cosh\eta_{\rm vir}-1)^2}{
                 (\sinh\eta_{\rm vir}-\eta_{\rm vir})^2}
              & (\Omega_0<1, ~ \lambda_0=0)
   \end{array}
 \right. ,
\end{equation}
where $\omega_{\rm vir}$ and $\eta_{\rm vir}$ are respectively given as
 $\omega_{\rm vir}\equiv1/\Omega_{\rm vir}-1$ and $ \eta_{\rm
 vir}\equiv\cosh^{-1}(2/\Omega_{\rm vir}-1)$, in terms of the density
 parameter at the collapse time, $\Omega_{\rm vir}$.

In practice, we focus on three specific profiles. (i) the original NFW
profile with $\alpha=1$ and $\nu=2$ (Navarro et al. 1997), (ii) the
modified NFW profile with $\alpha=\nu=3/2$ indicated by
higher-resolution simulations (Fukushige \& Makino 1997, 2001a,b; Moore
et al. 1998; Jing \& Suto 2000), (iii) the Hernquist profile with
$\alpha=1$ and $\nu=3$ (Hernquist 1990) for which the analytic
expression of the phase-space distribution function is known.

The two parameters $r_{\rm s}$ and $r_{\rm vir}$ are not independent,
and are related in terms of the concentration parameter:
\begin{equation}
  \label{eq: concentration}
c(M,z) \equiv \frac{r_{\rm vir}(M,z)}{r_\mathrm{s}(M,z)}.    
\end{equation}
The condition that the total mass inside $r_{\rm vir}$ is equal to $M$
relates $\delta_{\rm c}$ to $c$. Therefore the halo mass-dependence of
the above profiles is entirely specified by $c=c(M)$.
In the case of the original NFW profile, we use the 
approximate fitting function from the simulation data of
Bullock et al. (2001):
\begin{equation}
 c_{\rm B}(M)=\frac{8.0}{1+z}\,\,
 \left(\frac{M}{10^{14}\MO}\right)^{-0.13}. 
\label{eq: c_Bullock}
\end{equation}
For the other profiles, we first compute the amplitude of the two-point
correlation functions of dark matter following the procedure of Seljak
(2000) and Ma \& Fry (2000), and then find the amplitude of the
concentration parameter which reproduces the Peacock -- Dodds (1996)
fitting formula. This calibration yield $c (M)=c_{\rm B}(M)/2$ for the
modified NFW profile (Oguri et al. 2001), and $c(M)=c_{\rm B}(M)/3$ for
the Hernquist profile, which we adopt throughout the analysis below.

\subsection{Phase-space distribution function in a halo}

Our next task is to compute the phase-space distribution function in a
single halo from the given density profile (\ref{eq:univ}). While Sheth
(1996) and Sheth et al. (2001) simply adopt the Gaussian velocity
distribution function, we eliminate this assumption and derive the
velocity distribution function in a fully consistent manner.

For this purpose, we make use of the Jeans theorem which states that for
a spherically symmetric and stationary system, the solution of the
collisionless Boltzmann equation can be expressed as a function of the
specific binding energy, $E=\psi(r)-v^2/2$, alone.  Here we define
$\psi$ as the minus of the gravitational potential satisfying the
boundary condition of $\psi (r\to\infty)\to0$.

  One may wonder whether halos in hierarchical universes that should
experience repeated merger and destruction continually are well
approximated as stationary.  Nevertheless Natarajan, Hjorth \& van
Kampen (1997) and Hanyu \& Habe (2001) found that the phase-space
distribution function directly estimated from their particle simulations
agrees well with that derived from the Jeans theorem. Thus the above
assumption is justified, at least empirically.

\begin{figure}
\begin{center}
    \FigureFile(80mm,80mm){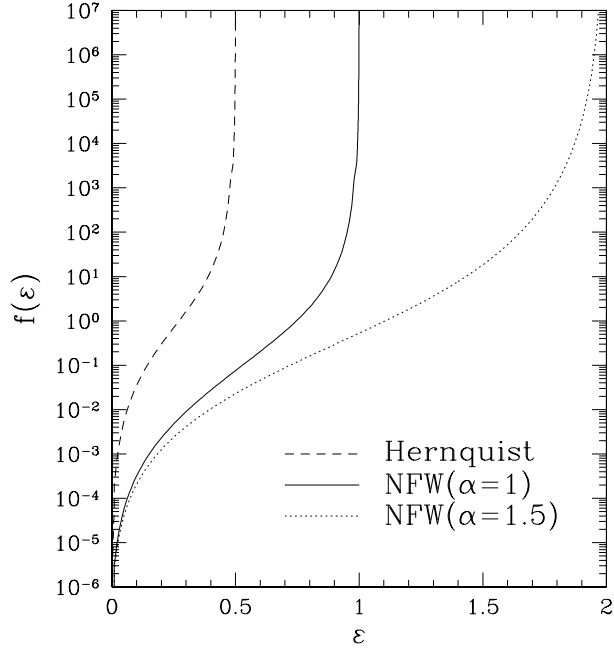}
\end{center}
\caption{The dimensionless phase-space distribution function
$f(\varepsilon)\equiv F(E|M)(Gm_s/r_s)^{3/2}/(\delta_c\bar{\rho})$ as a
function of dimensionless binding energy $\varepsilon= Er_s/(Gm_s)$;
NFW($\alpha=1$) profile ({\it solid}), NFW($\alpha=1.5$) profile ({\it
dotted}) and the Hernquist profile ({\it dashed}).  \label{fig:fe} }
\end{figure}

Then the phase-space distribution function $F(E|M)$ in a single halo is
directly computed from its given density profile $\rho(r|M)$ as follows
(e.g., Binney \& Tremaine 1987):
\begin{equation}
 F(E|M)=\frac{1}{\sqrt{8}\pi^2}\frac{d}{dE}
 \int_0^E\frac{d\rho}{d\psi}
 \frac{d\psi}{\sqrt{E-\psi}} .
\label{eq:abel}
\end{equation}
Figure \ref{fig:fe} plots the dimensionless phase-space distribution
function
 $f(\varepsilon)\equiv F(E|M)(Gm_s/r_s)^{3/2}/(\delta_c \bar\rho)$
evaluated numerically from equation (\ref{eq:abel}), where $m_s
\equiv4\pi r_s^3\delta_c\bar{\rho}$ is the characteristic mass of the halo
and $\varepsilon= Er_s/(Gm_s)$ is the dimensionless specific energy.
The Hernquist model has an analytical solution for $F(E|M)$ which is
reproduced by our numerical result
               almost within an accuracy of 2\%
               except for the tiny region $\varepsilon\sim0$,
               where the error reaches at 7\% but
               the effect is safely negligible for later analysis.
Since the three halo profiles that we adopt have a central cusp,
$f(\varepsilon)$ diverges at a corresponding value of $\varepsilon$.
The modified NFW profile has $f(\varepsilon)$ which extends more broadly
up to $\varepsilon \sim 2$ reflecting the stronger central concentration
than that of the original NFW case.

\subsection{Single-particle velocity distribution function in a halo}

\begin{figure}
\begin{center}
    \FigureFile(160mm,100mm){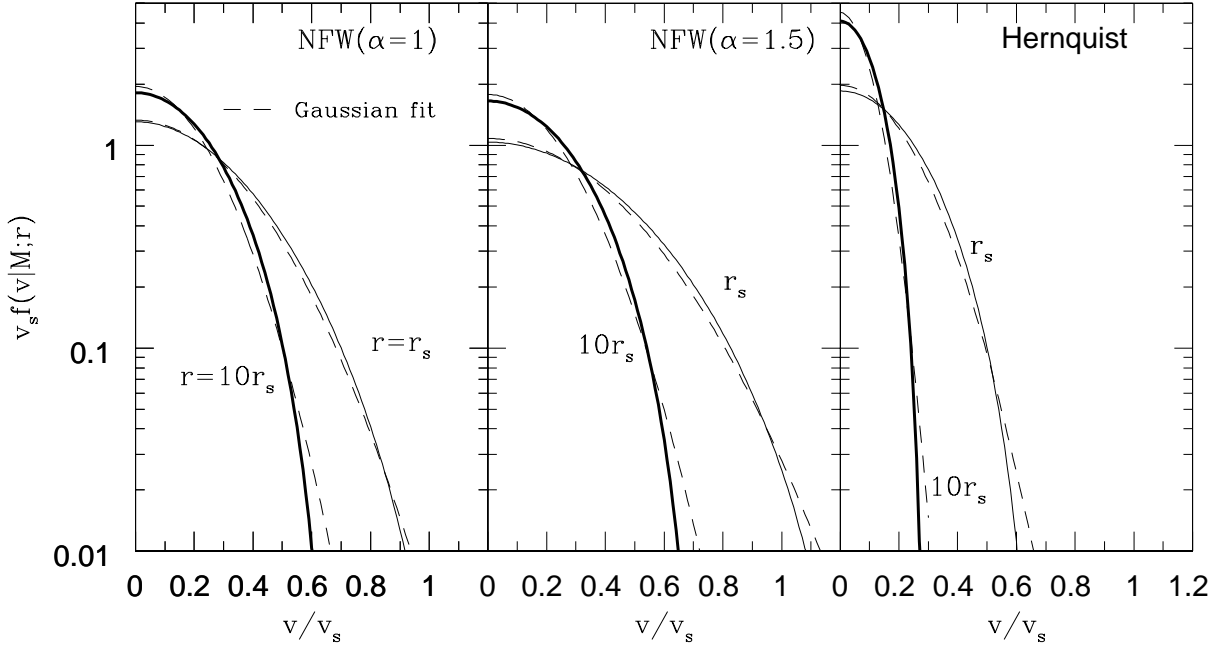}
\end{center}
\caption{Single-particle velocity distribution function in a single halo
at $r/r_{\rm s}=1$ ({\it thin-solid}) and $r/r_{\rm s}=10$ ({\it
thick-solid}). The dashed lines show the corresponding Gaussian fits
(see eq.[\ref{eq:f1v}]).  {\it Left}: NFW($\alpha=1$); {\it Middle}:
NFW($\alpha=1.5$); {\it Right}: Hernquist profile.  \label{fig:f1v} }
\end{figure}

Once the phase-space distribution function is given, one can also
compute one-dimensional single-particle velocity distribution function
along a particular direction by integrating over the other two components.
   Assuming the isotropic velocity distribution, one has
\begin{eqnarray}
 f_1(v_1|M;r) & \equiv & \frac{1}{\rho(r|M)}\,\,
 \int dv_2\int dv_3\,\,F(E|M) \nonumber\\
 &=& \frac{1}{\sqrt{2}\pi\,\,\rho(r|M)}\,\,\int_0^{E_1}dE\,\,
     \frac{d}{dE}\int_0^{E}\,\,
     \frac{d\rho}{d\psi}\frac{d\psi}{\sqrt{E-\psi}} \nonumber\\
 &=& \frac{1}{\sqrt{2}\pi\,\,\rho(r|M)}\,\,\int_0^{E_1}
     \frac{d\rho}{d\psi}\frac{d\psi}{\sqrt{E-\psi}},
\label{eq:f1}
\end{eqnarray}
where we project along the direction of $v_1$ and the quantity $E_{1}
\equiv \psi-v_1^{2}/2$ is the corresponding binding energy.  Figure
\ref{fig:f1v} shows the dimensionless velocity distribution function
$v_{\rm s}f_{1}(v|M;r)$ at $r/r_s=1$ ({\it thin lines}) and $r/r_s=10$
({\it thick lines}), where $v_{\rm s}\equiv (G m_s/r_s)^{1/2}$ is the
scaling velocity.  The figure indicates that the one-dimensional
velocity distribution function in a halo can be reasonably approximated
by the Gaussian:
\begin{equation}
 f_1(v|M;r) = \frac{1}{\sqrt{2\pi}\sigma(r|M)}
 \exp\left(-\frac{v^2}{2\sigma^{2}(r|M)}\right),
\label{eq:f1v}
\end{equation}
although it has a sharp cutoff around the escape velocity of the halo,
$v_{\rm esc}=(2\psi)^{1/2}$.  

In Figure \ref{fig:f1v}, the dashed lines indicate the Gaussian-fit
which has the same velocity dispersion evaluated from equation
(\ref{eq:f1}). It seems that the empirical Gaussian approximation can
reasonably reproduce the PVDF, and thus we use the approximation in the
numerical integrations below so as to reduce the computational time.

\begin{figure}
\begin{center}
    \FigureFile(80mm,80mm){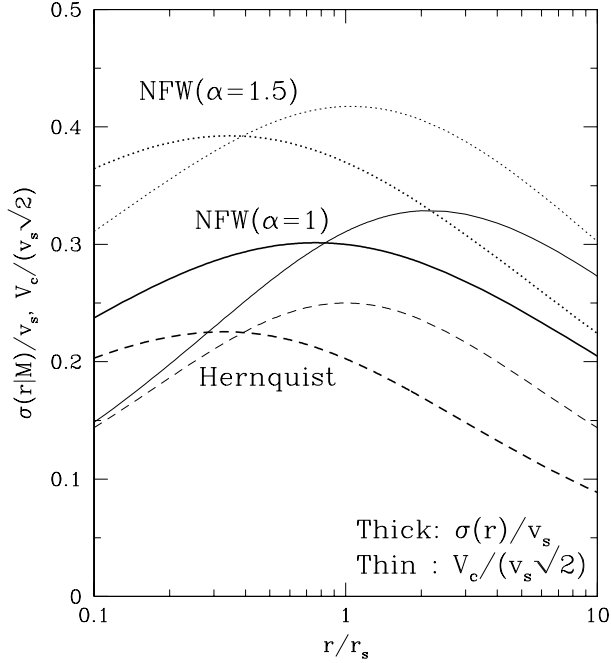}
\end{center}
\caption{The peculiar velocity dispersion $\sigma(r|M)$ of dark matter
particles in a single halo resulting from the Gaussian fit as a
function of position $r/r_{\rm s}$ ({\it thick lines}). 
For comparison, the thin lines show the circular velocity 
$V_{\rm c}$ evaluated from the relation $V_{\rm c}(r)=Gm(r)/r$, where 
$m(r)$ represents the mass inside the radius $r$. 
Solid line, dotted line, and dashed lines represent the results in 
NFW($\alpha=1$), NFW($\alpha=1.5$), and 
Hernquist profiles, respectively. \label{fig:svd} }
\end{figure}

Figure \ref{fig:svd} plots the velocity dispersion $\sigma(r|M)$
computed from the best-fit Gaussian, which clearly shows the
scale-dependence that was neglected in the previous analysis (Sheth
1996; Sheth et al. 2001).  Note also the different scale-dependence from
the circular velocity $V_{\rm c} \equiv Gm(r)/r$, where $m(r)$ is the
mass inside the radius $r$ (thin lines in Fig.\ref{fig:svd}).  In the
subsequent modeling of the PVDF, we use the Gaussian approximation with
the fitted $\sigma(r|M)$ rather than repeating the full numerical
integration.

\subsection{Pairwise relative velocity distribution function}

\begin{figure}
  \begin{center}
      \FigureFile(80mm,80mm){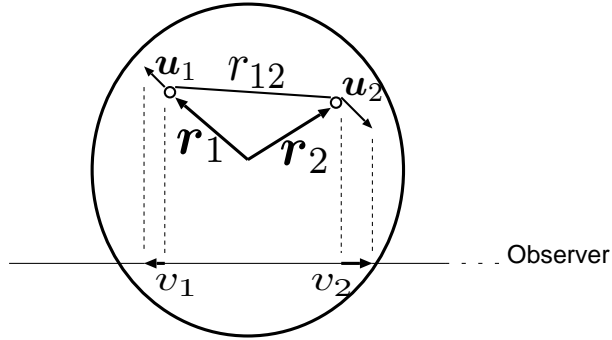}
  \end{center}
   \caption{The schematic picture of the relations of the relative
          positions and velocities
          used in the expressions (\ref{eq:1pvdf}) and (\ref{eq:normal}).
          The large circle and the two small circles denote a halo and
           a pair of particles, respectively. 
         The separation length of the pair $r_{12}$ is defined by
          $|\boldsymbol{r}_1-\boldsymbol{r}_2|$ and
         the relative pairwise velocity $v_{12}$ is defined by
          $v_1-v_2$.
          The quantities $v_1$ and $v_2$ are the line-of-sight component of
          the three dimensional velocities
           $\boldsymbol{u}_1$ and $\boldsymbol{u}_2$, respectively.
}
 \label{fig:r12}
\end{figure}

Finally we are in a position to estimate the PVDF combining the above
results.  Since we are interested in small scales, the particle pairs
with the corresponding separations are approximated to reside in the
common halo. Then Sheth (1996) derived the following expression for the
PVDF:
\begin{eqnarray}
&& f(v_{12};r_{12}) = {\cal N}^{-1}
  \int dM\,\,n(M)\,\int d^3\boldsymbol{r}_1d^3\boldsymbol{r}_2
  \,\,\rho(r_1|M)\rho(r_2|M)
\nonumber \\ 
&& ~~~~~~~ \times \int dv_1dv_2\,\,f_1(v_1|M;r_1)f_1(v_2|M;r_2)\,\,
   \delta_D(r_{12}-|\boldsymbol{r}_1-\boldsymbol{r}_2|)\,\,
   \delta_D(v_{12}-v_1+v_2),
\label{eq:1pvdf}
\end{eqnarray}
where $r_{12}$ is the pair-separation and ${\cal N}$ is the
 normalization factor given by
\begin{equation}
 {\cal N} = 
  \int dM \,\,n(M)\,\,\int d^3\boldsymbol{r}_1d^3\boldsymbol{r}_2
   \,\,\rho(r_1|M)\rho(r_2|M)\,\,
   \delta_D(r_{12}-|\boldsymbol{r}_1-\boldsymbol{r}_2|).
\label{eq:normal}
\end{equation}
  The relation between the quantities used in the expressions
  (\ref{eq:1pvdf}) and (\ref{eq:normal}) are schematically summarized in
  Figure \ref{fig:r12}.

While Sheth (1996) adopted the singular isothermal sphere
$\rho(r)\propto r^{-2}$ and therefore the Maxwellian for $f_{1}(v|M;r)$
with $r$-independent velocity dispersion, we are able to evaluate
equation (\ref{eq:1pvdf}) in more realistic situations as described in
the preceding subsections.

Applying the Gaussian fit (eq.[\ref{eq:f1v}]) for $f_{1}(v|M;r)$,
equation (\ref{eq:1pvdf}) is rewritten as
\begin{eqnarray}
\label{eq:pvdf1}
&& f_{12}(v_{12}|r_{12})
= {\cal N}^{-1}\,\,
\int dM\,\, n(M)\,\,\int d^{3}\boldsymbol{r}_1 \int d^{3}\boldsymbol{r}_2
\,\,\,\,\frac{\,\,\rho(r_1|M)\,\,\rho(r_2|M)\,\,}
{2\pi \sigma(r_1|M)\sigma(r_2|M)}
\nonumber \\
&&\times \int dv_1dv_2\,\, 
\exp\left[-\frac{v_1^2}{2\sigma^2(r_1|M)}-\frac{v_2^2}{2\sigma^2(r_2|M)}
  \right]
\delta_D(r_{12}-|\boldsymbol{r_1}-\boldsymbol{r_2}|)\,\,
 \delta_D(v_{12}-v_1+v_2).
\end{eqnarray}
The integrals over the two velocity components $v_1$ and $v_2$, and also
over the directions of the position vectors can be performed
analytically, and equation (\ref{eq:pvdf1}) reduces to
\begin{eqnarray}
\hspace*{-1.2cm}
f_{12}(v_{12}|r_{12}) &=&
\frac{\tilde{\cal N}^{-1}}{4\pi r_{12}^2\bar{\rho}^2}
  \,\,\int_{M_{\rm min}(r_{12})}^\infty dM\,\,n(M) \,\,
  r_{12}\int_{\max(0,r_{12}-r_{\rm s}c)}^{r_{\rm s}c}dr_2
  \int_{|r_{12}-r_2|}^{\min(r_{\rm s}c,r_{12}+r_2)}dr_1
  \nonumber\\
 &\times&
    r_1r_2\frac{\rho(r_1|M)\,\,\rho(r_2|M)}{
    \sqrt{2\pi\{\sigma^2(r_1|M)+\sigma^2(r_2|M)\}}}
\exp\left[-\frac{v_{12}^2}{2\{\sigma^2(r_1|M)+\sigma^2(r_2|M)\}}\right] ,
\label{eq:pvdf}
\end{eqnarray}
where $M_{\rm min}(r_{12})$ is the minimum mass of the halo including
the pair with separation $r_{12}$ (i.e., $r_{\rm vir}(M_{\rm min}) >
r_{12}/2$), and the normalization factor $\tilde{\cal N}$ is now
given by
\begin{eqnarray}
\tilde{\cal N}
 = \frac{1}{4\pi r_{12}^2\bar{\rho}^2}\int_{M_{\min}(r_{12})}^\infty 
 \hspace{-4ex}
   dM
 \,\,n(M)\,\,r_{12}\int_{\max(0,r_{12}-r_{\rm s}c)}^{r_{\rm s}c}
   \hspace{-3ex}
   dr_2
      \int_{|r_{12}-r_2|}^{\min(r_{\rm s}c,r_{12}+r_2)}
   \hspace{-5ex}
   dr_1
    r_1r_2\rho(r_1|M)\rho(r_2|M) .
\label{eq:xi_1h}
\end{eqnarray}
Actually it turns out that this term corresponds to the one-halo
contribution of the two-point correlation function, $\xi_{\rm
1h}(r_{12})$, in the dark halo approach (Seljak 2000, Ma \& Fry 2000).

\begin{figure}
\begin{center}
    \FigureFile(160mm,110mm){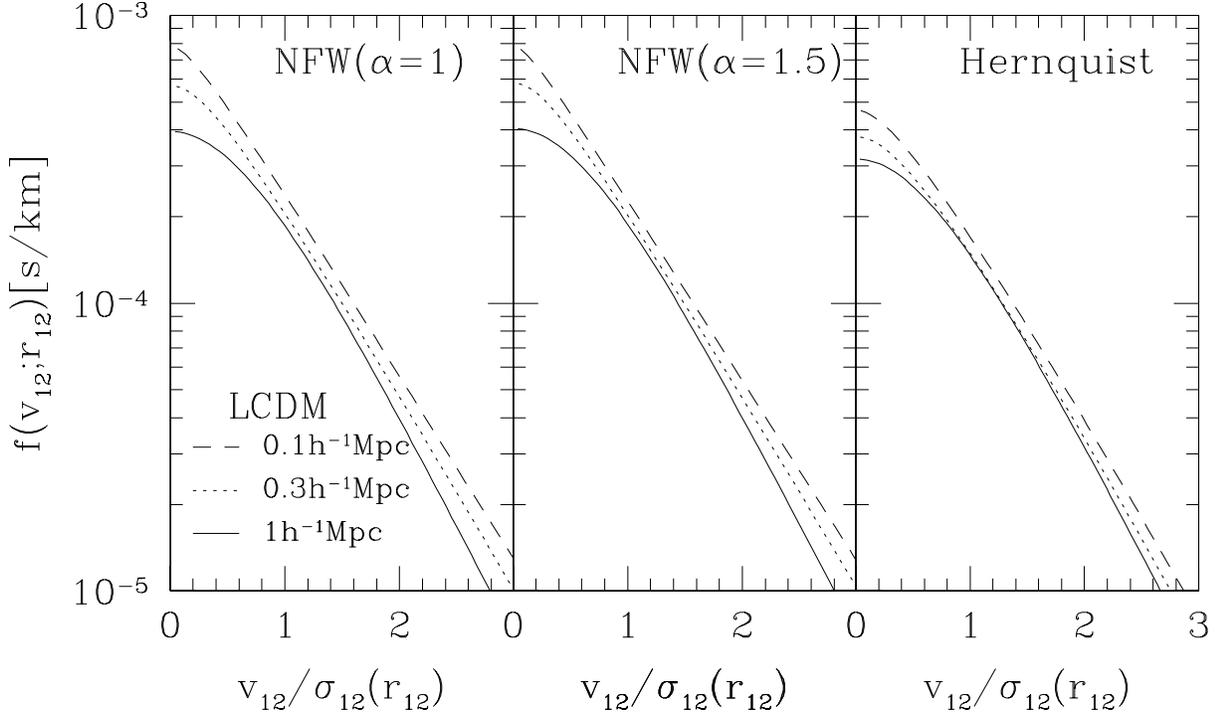}
\end{center}
\caption{Pairwise velocity distribution function averaged over the halo
 mass function in the LCDM universe.  {\it Left}: NFW($\alpha=1$); {\it
 Middle}: NFW($\alpha=1.5$); {\it Right}: Hernquist profile.  Solid,
 dotted and dashed lines indicate the results at the pair separation of
 $1h^{-1}$Mpc, $0.3h^{-1}$Mpc, and $0.1h^{-1}$Mpc.  } \label{fig:f2v}
\end{figure}

Figure \ref{fig:f2v} plots the resulting PVDF (in the LCDM model) at
$r_{12}=1h^{-1}$, $0.3h^{-1}$ and $0.1h^{-1}$Mpc against the pairwise
velocity $v_{12}$ normalized by the PVD at each separation.  We adopt
the Press-Schechter mass function for definiteness.
  Note that the quantitatively similar behavior is obtained for
  other cosmological models, but with different PVDs.
 The PVDF for small
separation pairs ($r_{12} \ll 1h^{-1}$Mpc) is well described by the
exponential distribution. As $r_{12}$ increases, the central region
resembles the Gaussian distribution while the exponential tail is still
clear at large velocities.  Neither the inner nor outer slope of the
density profile produces any systematic difference in the non-Gaussian
tails of PVDF, in contrast to the single-particle velocity distribution
in Figure \ref{fig:f1v}.  This qualitative behavior is in complete
agreement with the result of Sheth (1996) assuming the scale-free model
and the singular isothermal sphere.  Therefore we conclude that the
exponential distribution of the PVDF is a rather general consequence in
the gravitational instability picture fairly independent of the
underlying cosmological model.

\section{Pairwise relative peculiar velocity dispersions}

\subsection{Dark matter particles}

We have shown that the
   {\it shape}
 of the PVDF is well approximated by the
exponential in a fairly insensitive manner to either the cosmological
model or the dark halo density profile.
  Note, however, that this does not implies the PVD is determined independently
  of the cosmology, but rather it is the basic
  source for the cosmological model-dependence,
  through the mass function of halos.
 Since the PVDF is already
  obtained, one may evaluate the PVD $\sigma_{12}(r_{12})$ directly as
\begin{equation}
 \sigma^2_{12}(r_{12}) =\int_{-\infty}^\infty dv_{12}\,\,
 f_{12}(v_{12};r_{12})\,\, v_{12}^2. 
\label{eq:disp}
\end{equation}
We note, however, that the above expression is valid only when the
one-halo term $\xi_{\rm 1h}(r_{12})$ is sufficiently larger than unity.
If one takes account of particle pairs residing in two different halos
that we neglect in the present modeling, one should rather replace the
normalization factor $\tilde{N}$ by $1+\xi(r_{12})$ since the factor
physically corresponds to the relative probability of finding a pair
at separation $r_{12}$. This consideration implies the normalization of
the PVD from the one-halo contribution should be
\begin{equation}
 \sigma^2_{12}(r_{12}) = \frac{\xi_{\rm 1h}(r_{12})}{1+\xi(r_{12})} 
 \int_{-\infty}^\infty dv_{12}\,\,f_{12}(v_{12};r_{12})\,\, v_{12}^2.   
\label{eq:ndisp}
\end{equation}
In fact, this agrees with equation (21) of Sheth et al. (2001).  We
compute the two-point correlation function, $\xi(r_{12})=\xi_{\rm
1h}(r_{12})+\xi_{\rm 2h}(r_{12})$, on the basis of the dark halo
approach (Seljak 2000), and we also confirm that the resulting
$\xi(r_{12})$ is in good agreement with the fitting formula of Peacock
\& Dodds (1996) by an appropriate choice of the concentration parameter
$c(M)$ as discussed in subsection \ref{subsec:rho}.

\begin{figure}
\begin{center}
    \FigureFile(160mm,100mm){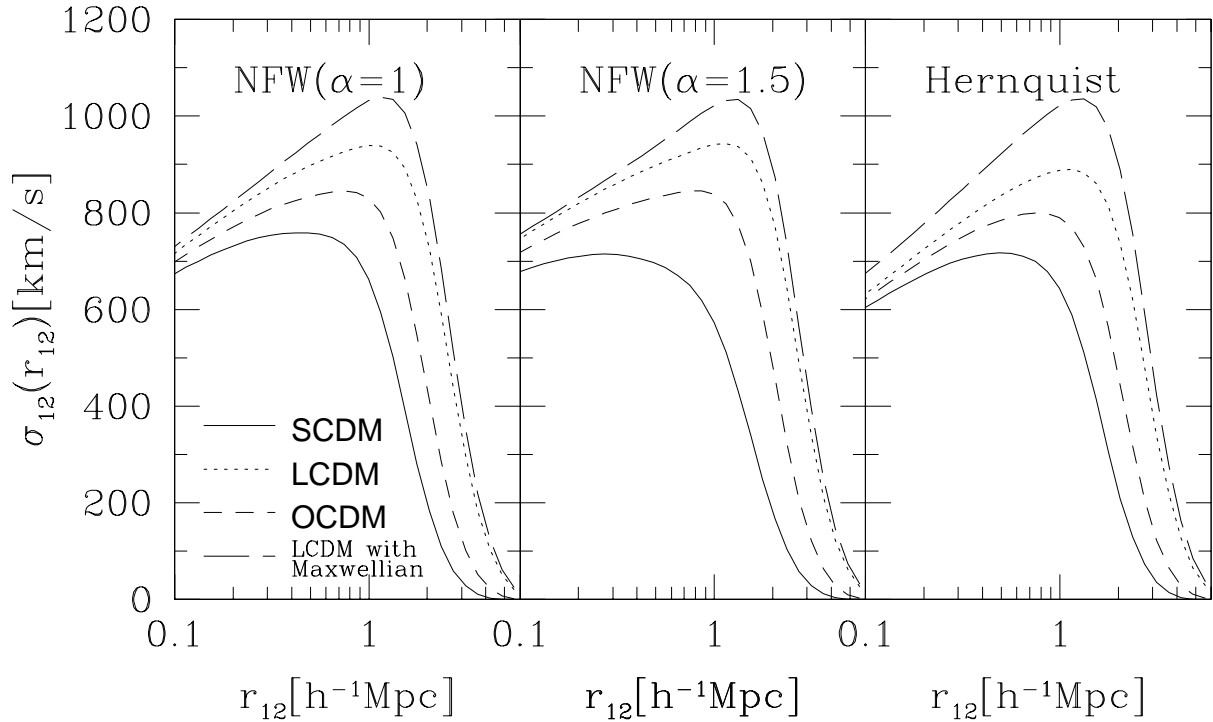}
\end{center}
\caption{The pairwise velocity dispersion of dark matter particles.
{\it Left}: NFW($\alpha=1$); {\it
 Middle}: NFW($\alpha=1.5$); {\it Right}: Hernquist profile. 
Solid, dotted and dashed lines indicate the results in
SCDM, LCDM,  and OCDM.  Long-dashed line shows the result in LCDM under
 the assumption that each halo 
        has the Maxwellian velocity distribution function
          with constant velocity dispersion.
}  \label{fig:lsigma}
\end{figure}

Figure \ref{fig:lsigma} shows the PVD calculated according to equation
(\ref{eq:ndisp}). The comparison among the different model predictions
indicates that the amplitude of PVD sensitively depends on the
cosmological parameters
        through the mass function $n(M)$,
 but is almost insensitive to the density
profile of dark halo.
  Note, however, that this is partly because we
have chosen the value of $c(M)$ so as to reproduce the same $\xi(r)$
irrespectively of the density profile.  We also show the result with
neglecting the scale-dependence of the velocity distribution in the case
of the LCDM model ({\it long-dashed} lines). This indicates that the
isothermal approximation (Sheth et al. 2001) is quite acceptable in 
predicting the PVD.

\begin{figure}
\begin{center}
    \FigureFile(160mm,100mm){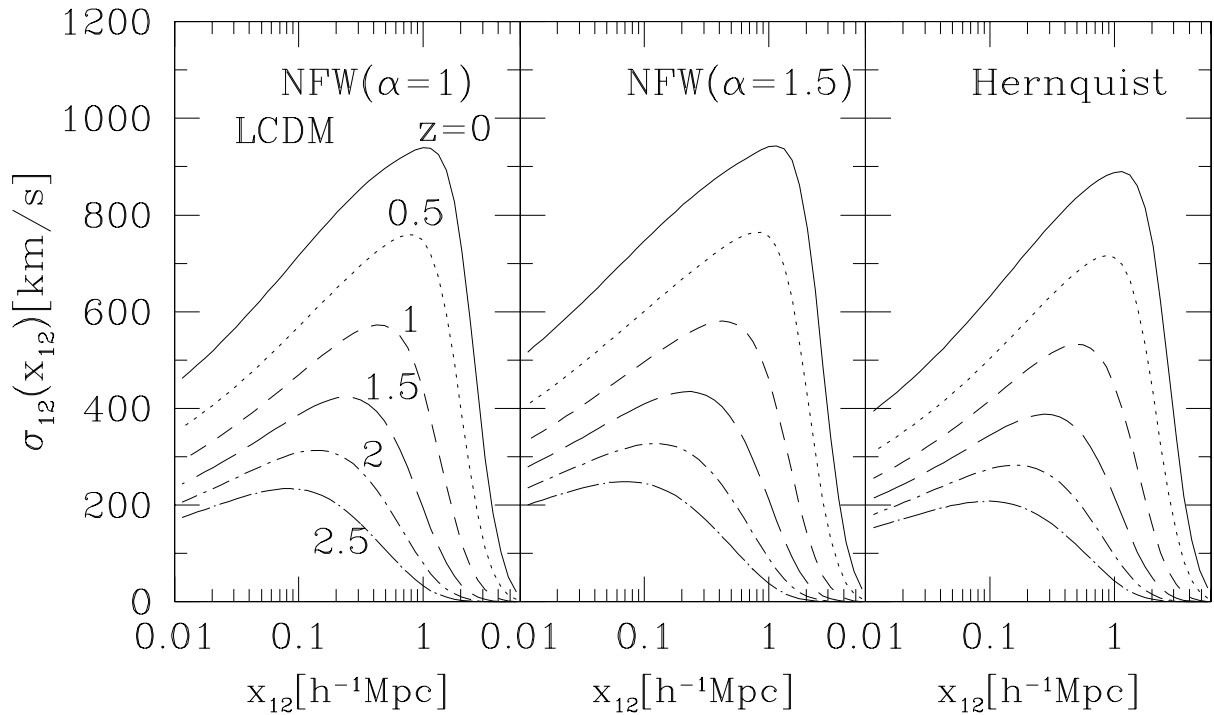}
\end{center}
\caption{The pairwise velocity dispersion of dark matter particles
in LCDM at $z=0$, 0.5, 1.0, 1.5, 2.0 and 2.5 from top to bottom.
{\it Left}: NFW($\alpha=1$); {\it
 Middle}: NFW($\alpha=1.5$); {\it Right}: Hernquist profile. 
}  \label{fig:glsigma}
\end{figure}

We attempted an empirical fitting to our numerical results at different
redshifts (Fig.\ref{fig:glsigma}) by adopting a power-law form:
\begin{eqnarray}
\label{eq:sigmafit}
\sigma_{12}(x_{12}) = A \left(\frac{x_{12}}{1h^{-1} {\rm Mpc}}\right)^p ,
\end{eqnarray}
in terms of the {\it comoving} pair separation $x_{12} \equiv
r_{12}(1+z)$.  The values of the amplitude $A$ and the power-law index
$p$ fitted for $0.01 h^{-1} {\rm Mpc} < x_{12} < x_{\rm max}$ are listed
in Table~\ref{tab:power-law-fit} in the LCDM model,
                where $x_{\rm max}$ is the comoving separation
                 at which the PVD becomes maximum.
                Since the two-halo term $\xi_{\rm 2h}(r_{12})$ becomes 
                 dominant contribution to $\xi(r_{12})$ for larger separations,
                 equation (\ref{eq:sigmafit}) becomes 
                  inaccurate for $x> x_{\rm max}$. 	
                       The fit is accurate within 5 \%, 
                       which is comparable to the other systematic errors
                       including the numerical integration or the
                       Gaussian approximation.
 This result may be compared with the independent prediction
based on the cosmic virial theorem (Peebles 1976; Suto 1993; Suto \&
Jing 1997). For reference, if the correlation function of dark matter is
given by $(r/5.4h^{-1} {\rm Mpc})^{-1.8}$, the cosmic virial theorem
implies that
\begin{eqnarray}
\label{eq:sigmacvt}
 v_{12} (r_{12})_{\rm \scriptscriptstyle CVT} = 990
\left(\frac{\Omega_0}{0.3}\right)^{1/2}
\left(\frac{Q}{2.0}\right)^{1/2}
 \left({r_{12} \over 1h^{-1} {\rm Mpc} } \right)^{0.1} {\rm km/s} ,
\end{eqnarray}
where $Q$ is the normalized amplitude of the three-point correlation
function. The value of $Q$ for dark matter clustering is somewhat
uncertain, but may be close to $2$ in LCDM (Suto 1993), and thus this
estimate is fairly consistent with our prediction presented here.

\begin{table}[thb]
\caption{Power-law fits to the relative peculiar velocity dispersion in
 the LCDM model.    \label{tab:power-law-fit}}
\begin{center}
 \begin{tabular}{cc|cc|cc|cc}
      \multicolumn{2}{c}{}&
      \multicolumn{2}{|c|}{NFW($\alpha=1$)} &
      \multicolumn{2}{|c|}{NFW($\alpha=1.5$)} &
      \multicolumn{2}{|c}{Hernquist}\\ \hline 
      \multicolumn{1}{c}{$z$}&
      \multicolumn{1}{c}{$x_{\max}$[$h^{-1}$Mpc]}&
      \multicolumn{1}{|c}{$A$[km/s]}&
      \multicolumn{1}{c|}{$p$}&
      \multicolumn{1}{|c}{$A$[km/s]}&
      \multicolumn{1}{c|}{$p$}&
      \multicolumn{1}{|c}{$A$[km/s]}&
      \multicolumn{1}{c}{$p$} \\ \hline 
0.0 & 1.0  & 1016 & 0.17 & 997 & 0.14 & 957 & 0.19 \\
0.5 & 0.9  & 846  & 0.18 & 834 & 0.15 & 791 & 0.20 \\
1.0 & 0.4  & 731  & 0.20 & 713 & 0.16 & 675 & 0.22\\
1.5 & 0.25 & 578  & 0.19 & 517 & 0.15 & 533 & 0.20\\
2.0 & 0.15 & 458  & 0.17 & 442 & 0.13 & 416 & 0.18\\
2.5 & 0.11 & 339  & 0.14 & 325 & 0.10 & 304 & 0.15\\
    \end{tabular}
  \end{center}
\end{table}

\subsection{Effect of biasing}
\label{sec:bias}

It is well known that the PVD of the observed galaxies is generally
smaller than the value predicted in current popular models
 (e.g., Davis \&
Peebles 1983; Mo, Jing \& B\"orner 1993; Suto 1993; Suto \& Jing 1997).
This may be interpreted as a manifestation of the spatial biasing of
galaxies relative to dark matter.

Jing, Mo \& B\"orner (1998) analyzed the Las Campanas Redshift Survey
galaxies and developed a phenomenological biasing model, CLuster
underWeight bias (CLW, hereafter) which successfully accounts for the
amplitudes of the two-point correlation function and the PVD
simultaneously.  More recently Jing, B\"orner \& Suto (2002) performed
the similar analysis of galaxies in the PSCz catalog (Saunders et
al. 2000) which are selected from the InfraRed Astronomical Satellite
(IRAS) Point Source Catalog (PSC; Beichman et al. 1988). They found that
the IRAS selected galaxies, which are likely to be dominated by
late-types, have significantly smaller PVD than those in other
catalogues. In addition, they applied the CLW bias scheme to mock
samples from $N$-body simulations and concluded that the PVD of the PSCz
galaxies is significantly smaller than those predicted from the CLW bias
in the popular CDM models.  In this subsection, we revisit this issue
combing the biasing effect with our analytical model of the PVD in a
complementary manner to the direct method using the $N$-body data.

The CLW scheme can be applied to our analytical model easily.  According
to Jing et al. (2002), we adopt that the selection function of galaxies
in a halo of mass $M$ is proportional to $(M/10^{14}\MO)^{-\beta}$. This
is simply equivalent to replacing the mass function $n(M)$ in equation
(\ref{eq:pvdf}) by $n(M)(M/10^{14}\MO)^{-\beta}$.  Because this biasing
model puts lower weight on the massive clusters where the velocity
dispersion is large, increasing $\beta$ suppresses the mean PVD.  In
fact, the Las Campanas redshift survey data are consistent with
$\beta=0.08$, and the PSCz data prefer a much larger value $\beta=0.25$.
Physically speaking, this phenomenological prescription should be
understood as the dependence of the efficiency of galaxy formation on
the mass of the hosting halo.

In addition we consider another biasing to take into account the
observed density-morphology relation of galaxies. Since spiral galaxies
preferentially avoid the central region of massive clusters (i.e., halos
in the present context), the PVD of spirals is generally suppressed
especially in the case of the modified NFW profile that has stronger
central concentration (c.f., Fig.\ref{fig:svd}).  We attempt to
incorporate this effect by introducing the selection probability that
depends on a distance from the center of the halo:
\begin{equation}
\label{eq:dmr}
 p(r|M)=a+\frac{r}{r_{\rm vir}(M)}b .
\end{equation}
We set $a=0.2$ and $b=0.6$ so as to reproduce the observed ratio of
spirals to ellipticals; 2:8 in the inner part and 8:2 in the outer part.
In practice, we calculate the PVD by replacing $\rho(r|M)$ by
$\rho(r|M)p(r|M)$ in equation (\ref{eq:ndisp}).

\begin{figure}
\begin{center}
    \FigureFile(80mm,80mm){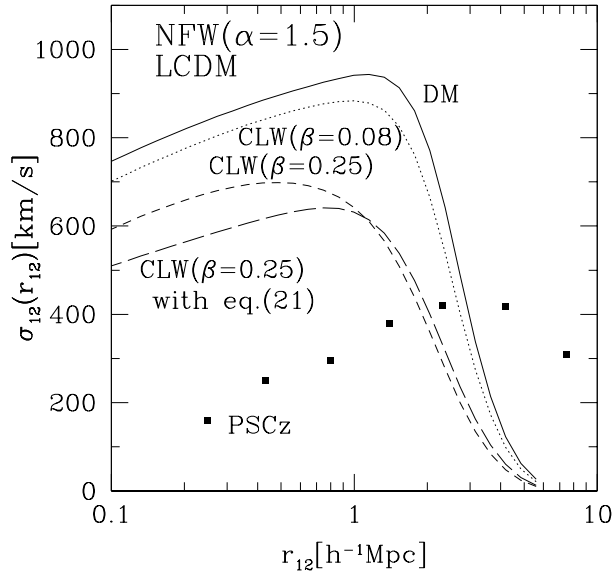}
\end{center}
\caption{Pairwise peculiar velocity dispersion for {\it galaxies} with
 empirical biasing scheme in LCDM; {\it Solid}: dark matter particles.
{\it Dotted}: CLW with $\beta=0.08$,
{\it Short-dashed}: CLW with $\beta=0.25$,
{\it Long-dashed}: CLW with $\beta=0.25$ and density-morphology
 relation.
The filled squares indicate the values for the PSCz galaxies estimated
 by Jing et al. (2002).}
\label{fig:bias}
\end{figure}

The results are shown in Figure \ref{fig:bias} for the modified NFW
profile in the LCDM model. The dotted and short-dashed lines represent
the results taking account of the CLW bias effect while the long-dashed
lines consider the density--morphology relation (eq.[\ref{eq:dmr}]) in
addition to the CLW bias. The degree of suppression of the PVD is in
agreement with the simulation results of Jing et al. (2002).  Even with
the density--morphology relation, the PVD of the IRAS PSCz galaxies is
too small to be reconciled in the current model as Jing et al. (2002)
claimed.

\section{Summary}

We have presented a detailed prediction for the pairwise peculiar
velocity distribution function (PVDF) applying the dark matter halo
approach. In particular, we have derived the PVDF in a direct and
self-consistent manner with the assumed density profile for dark matter
halo for the first time.  On the other hand, we neglect the halo-halo
contribution assuming that any pair of particles resides in a common
halo. Thus our predictions are quantitatively valid only on small scales
$\ltsim 1h^{-1}$Mpc, but our result turns out to be fairly close to the
previous one by Sheth (1996) and Sheth et al. (2001) who assumed the
{\it isothermal} velocity distribution in a single halo. In this sense,
our independent approach may be regarded as to provide an empirical
justification for their simplifying assumptions, and also our
predictions are fairly accurate on those small scales.

We have shown that the shape of the PVDF is well approximated by the
exponential in a fairly insensitive manner to either the cosmological
model or the dark halo density profile. The dependence on the PVD on the
halo density profiles that we employed is also fairly small, yielding
the difference less than about 10 percent.

We have also obtained a practical fitting formula for the PVD of dark
matter particles (eq.[\ref{eq:sigmafit}]) at different cosmological
models as a function of the pair separation. This may be useful in
modeling redshift-space distortion of clustering. The result is in
reasonable agreement with the estimate on the basis of the cosmic virial
theorem. Furthermore we apply an empirical biasing scheme into our
model and attempt to predict the PVD of {\it galaxies}. We can reproduce
the previous simulation results on the basis of our analytical method,
and also confirmed that the very small PVD estimated for the PSCz
galaxies (Jing et al. 2002) is difficult to be reconciled with a
simplistic biasing model and/or the underlying CDM model.

  The discrepancy between the prediction and the observation shown in
  section \ref{sec:bias} indicates the presence of velocity bias,
  in addition to the other selection effects.
  In fact, each luminous galaxy is a clump composed of the baryon
  as well as dark matter particles,
  whose bulk motion might not trace the random motion of the individual
  dark matter particles.
  In this case, the galaxy-galaxy interaction through tidal field or
  gas pressure of baryon might be an important source for velocity bias.
  At least, our present treatment using dark matter halo approach can 
  provide a quantitative prediction for the PVD of the dark matter 
  particles. Therefore next step, the effect of velocity bias including these 
  interactions should be incorporated into our scheme to account for 
  the PVD of the galaxies.  

\vspace*{1cm}

T. K. thanks Chiaki Hikage, Issha Kayo, and Masamune Oguri for
discussions and comments on the manuscript.  This research was supported
in part by Monbu-Kagakusho (07CE2002, 12640231).


%
\end{document}